\documentclass[aps,twocolumn,superscriptaddress,nofootinbib,a4paper,longbibliography]{revtex4-2}

\usepackage{times}
\usepackage{epsfig}
\usepackage{amsfonts}
\usepackage{amsmath}
\usepackage{amsthm}
\usepackage{amssymb}					
\usepackage{amsthm}
\usepackage{dsfont}
\usepackage{bm}
\usepackage{mathtools}
\usepackage{color}
\usepackage{multirow}
\usepackage[normalem]{ulem}
\newcommand{\stkout}[1]{\ifmmode\text{\sout{\ensuremath{#1}}}\else\sout{#1}\fi}
\usepackage{latexsym}
\usepackage{mathrsfs}
\usepackage{natbib}
\usepackage{verbatim}
\usepackage[T1]{fontenc}
\usepackage{float}
\usepackage{graphicx}
\usepackage{subcaption}
\usepackage{physics}
\usepackage{soul}
\usepackage{caption}
\usepackage{subcaption}
\usepackage{diagbox}
\usepackage{bbm}
\usepackage{tablefootnote}
\usepackage{adjustbox}
\usepackage[dvipsnames]{xcolor}
\usepackage[font=small,labelfont=bf]{caption}

\usepackage{xspace}

\theoremstyle{definition}

\newcommand{\blue}[1]{\textcolor{blue}{#1}}

\newcommand{\bracket}[3]{\langle#1|#2|#3\rangle}

\usepackage{amssymb}
\usepackage{amsthm}
\usepackage{mathtools} 
\usepackage[customcolors]{hf-tikz} 
\usetikzlibrary{patterns}
\usetikzlibrary{matrix,decorations.pathreplacing}

\pgfkeys{tikz/mymatrixenv/.style={decoration={brace},every left delimiter/.style={xshift=8pt},every right delimiter/.style={xshift=-8pt}}}
\pgfkeys{tikz/mymatrix/.style={matrix of math nodes,nodes in empty cells,left delimiter={[},right delimiter={]},inner sep=1pt,outer sep=1.5pt,column sep=2pt,row sep=2pt,nodes={minimum width=20pt,minimum height=10pt,anchor=center,inner sep=0pt,outer sep=0pt}}}
\pgfkeys{tikz/mymatrixbrace/.style={decorate,thick}}

\usepackage[colorlinks=true,linkcolor=blue,citecolor=magenta,urlcolor=blue]{hyperref}

\begin{document}
	

\title{Semidefinite relaxations for high-dimensional entanglement in the steering scenario}
	
\author{Nicola D'Alessandro}
\affiliation{Physics Department and NanoLund, Lund University, Box 118, 22100 Lund, Sweden.}
		
\author{Carles Roch i Carceller}	
\affiliation{Physics Department and NanoLund, Lund University, Box 118, 22100 Lund, Sweden.}
	
\author{Armin Tavakoli}
\affiliation{Physics Department and NanoLund, Lund University, Box 118, 22100 Lund, Sweden.}
	
\begin{abstract}
We introduce semidefinite programming hierarchies for benchmarking relevant entanglement properties in the high-dimensional steering scenario. Firstly, we provide a general method for detecting the entanglement dimensionality through certification of the Schmidt number. Its key feature is that the computational cost is independent of the Schmidt number under consideration. Secondly, we provide a method to estimate the fidelity of the source with any maximally entangled state. Using only basic computational means, we demonstrate the usefulness of these methods, which can be directly used to analyse experiments on high-dimensional systems. 
\end{abstract}
	
	\date{\today}

	\maketitle

\textit{Introduction.---} Quantum technologies greatly benefit from correlations generated between entangled particles. While standard entanglement involves qubits, it is well-known that two entangled particles with a higher-than-qubit local dimension can exhibit stronger correlations. This is not only of conceptual interest but it also enables improved noise- and loss-tolerance in quantum key distribution \cite{Cerf2002} and steering experiments \cite{Skrzypczyk2015, Marciniak2015}, and allows for reduced detection efficiencies in Bell inequality tests \cite{Vertesi2010}.  Therefore, much research has been focused on certifying the dimension of entangled states; see e.g.~the recent experiments \cite{Krenn2014, Erhard2018, Bavaresco2018, Herrera2020, Hu2021, Goel2024}.

\begin{figure}
	\centering
	\includegraphics[width=1\columnwidth]{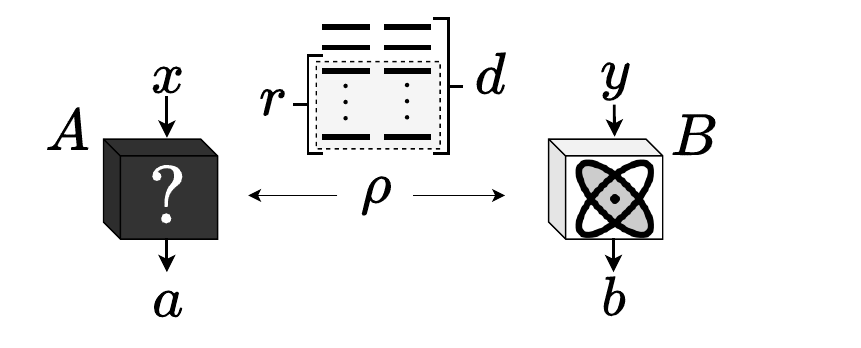}
	\caption{Steering scenario. A state $\rho$ is shared between two parties Alice ($A$) and Bob ($B$), who have inputs $(x,y)$ and outputs $(a,b)$. Alice's measurements are uncharacterised (black box) while Bob's measurements are known (white box). From  the local states of Bob, we bound the Schmidt number, $r$, of the shared state and its fidelity with the maximally entangled state in dimension $d$.}
	\label{fig:steering}
\end{figure}

Many methods and criteria have been developed to detect the entanglement dimensionality (a.k.a the Schmidt number \cite{Terhal2000})  of bipartite entangled states \cite{Bavaresco2018, Weilenmann2020, Morelli2023, Wyderka2023, Liu2024, Tavakoli2024}. This endevour has naturally been extended to more demanding scenarios, in which not all quantum operations are assumed to be characterised. A well-known example is the steering scenario \cite{Uola2020}, in which one party's (Alice's) operations are uncharacterised and entanglement therefore must be certified only from the local states observed by the other party, Bob (see Fig.~\ref{fig:steering}). This scenario has been studied for both fundamental interest and for practical interest as a platform for one-sided device-independent quantum information protocols \cite{Cavalcanti_2017}.

High-dimensional entanglement leads to significant correlation advantages for steering and it has therefore been the core of many steering experiments \cite{Zeng2018, Li2018, Wang2018, Srivastav2022, Huang2021, QuPRL2022, Qu22b}. However, in order to show that observed correlations are genuinely due to high dimensionality, one must falsify the hypothesis that the local states of Bob could have been generated in a quantum model that uses only lower-dimensional entanglement \cite{SebSteering2021}.  Analytical criteria for certifying the Schmidt number in the steering scenario have been put forward in  \cite{SebSteering2021} but these criteria are not tight and they are limited to a specific correlation test using two measurement bases. It is known that quantum steering greatly benefits from the use of many measurement bases \cite{Saunders2010, Bennet2012}, but current analytical criteria for certifying the Schmidt number from many bases can only slightly capitalize on this potential \cite{Designolle2022}. Recently, it has been proposed to address the problem via a semidefinite programming (SDP) approach \cite{CompleteHierarchy}. While this approach is general, it is less practical because its computational cost scales rapidly in the hypothesised Schmidt number. 

Here, we develop efficient SDP relaxation hierarchies for quantifying high-dimensional entanglement in the steering scenario; both in terms of the Schmidt number and in terms of fidelity with the maximally entangled state. Both of these are standard quality benchmarks in experiments. For the Schmidt number, our SDP method contrasts that of \cite{CompleteHierarchy} since its computational requirements have no dependence on the hypothesised Schmidt number and it is therefore scalable in this parameter. This allows us both to compute relevant steering bounds in scenarios that were previously out of reach for SDP methods and to improve previously known results. For  bounding the entanglement fidelity, we are not aware of alternative SDP methods. Through  case studies, we show that our method can closely approximate the exact fidelity results. Taken together, these methods enable versatile and pratical entanglement quantification in the high-dimensional steering scenario.

\textit{Preliminaries.---} Consider that Alice and Bob share an unknown bipartite quantum state $\rho_{\text{AB}}$. Alice selects an input $x$ and performs an unknown measurement $\{A_{a|x}\}$ on her part, where $a$ denotes the outcome. After Alice measures, the un-normalised post-measurement states in Bob's side become
\begin{equation}
\label{eq:sigma}
\sigma_{a|x}=\tr_\text{A}\left[(A_{a|x}\otimes\openone_\text{B})\rho_{\text{AB}}\right] \ ,
\end{equation}
where the local probability distribution in Alice's side is given by $p(a|x)=\tr(\sigma_{a|x})$. The set $\bm\sigma\equiv \{\sigma_{a|x}\}$  is called the assemblage, and is characterised by positivity $\sigma_{a|x}\succeq 0$ $\forall a,x$,  no-signaling $\sum_a \sigma_{a|x}=\tau$ $\forall x$, and normalisation $\tr(\tau)=1$, for $\tau$ being an arbitrary quantum state. We say that $\bm{\sigma}$ is steerable if it cannot be generated by any set of local measurements on any separable state. Although this can conveniently be decided by  SDP \cite{Cavalcanti_2017}, it requires Bob to perform full tomography on the assemblage. Alternatively, one can base steering tests on witness quantities, in which Bob performs only a smaller set of measurements $\{B_{b|y}\}$ where $y$ is the input and $b$ the outcome. Witness tests take the form
\begin{equation}\label{witness}
	W=\sum_{a,b,x,y}c_{abxy}\tr(\sigma_{a|x}B_{b|y})\leq \beta \ ,
\end{equation}
for some real coefficients $c_{abxy}$. These inequalities are respected by all unsteerable assemblages but can be violated otherwise. Such criteria are well-known and standard \cite{Cavalcanti_2017}.

Here, we are interested in a more challenging task: using $\bm\sigma$, Alice must convince Bob that $\rho_\text{AB}$ has Schmidt number at least $r$. The Schmidt number is the smallest local dimension necessary to generate $\rho_\text{AB}$ via a mixture of pure states \cite{Terhal2000}. Formally, it is written
\begin{align}\nonumber
	r(\rho_\text{AB})\equiv \min_{\{q_i\},\{\psi_i\}}  \Big\{&r_\text{max}: \quad  \rho_\text{AB}=\sum_i q_i \ketbra{\psi_i}\\\label{SN}
	&\text{and} \quad r_\text{max}=\max_i \rank(\psi_i^\text{A})\Big\} \ ,
\end{align}
where $\{q_i\}_i$ is a probability distribution and $\psi_i^\text{A}=\tr_\text{B}\left(\ketbra{\psi_i}\right)$. We denote by $\mathcal{S}_r$ the set of assemblages \eqref{eq:sigma} compatible with local measurements on states of Schmidt number at most $r$. Notice that choosing $r=1$ ($\rho_{\text{AB}}$ separable) reduces the membership problem for $\mathcal{S}_r$ to standard steering detection. 
However, when $r>1$ there is no longer a simple and general procedure for deciding whether  $\bm\sigma\in\mathcal{S}_r$.

The Schmidt number reveals the genuine entanglement dimensionality of $\rho_{\text{AB}}$ but it does not necessarily determine how accurately an entanglement preparation device works. It is therefore natural to also consider using $\bm{\sigma}$ to bound the fidelity between $\rho_{\text{AB}}$ and the desired target state produced by the source.  The most natural target state is the maximally entangled state, which up to local unitaries is given by $\ket{\phi^+_d}=\frac{1}{\sqrt{d}}\sum_{i=0}^{d-1} \ket{ii}$. The optimised fidelity is written
\begin{equation}\label{fidelity}
	F(\rho_{\text{AB}})=\max_{\Lambda} \ \ \ \bracket{\phi^+_d}{(\Lambda\otimes\openone)[\rho_{\text{AB}}]}{\phi^+_d} \ ,
\end{equation} 
where $\Lambda$ is some local extraction channel in Alice's lab. We refer to \eqref{fidelity} as the entanglement fidelity. We will estimate it both from both given assemblages \eqref{eq:sigma} and witness values  \eqref{witness}. 

\textit{Hierarchy for Schmidt number certification.---}
Consider that Alice and Bob share a pure state $\rho_{\text{AB}}=\ketbra{\psi}$ where Bob's dimension is $d$. If the Schmidt number of the state is at most $r\leq d$, the reduced state $\rho_\text{A}=\tr_\text{B}(\rho_{\text{AB}})$ is fully supported on an $r$-dimensional subspace. We call the projector onto this subspace $\Pi$.  W.l.g~we can restrict Alice's measurements $\{A_{a|x}\}$ to this support space of $\rho_\text{A}$, and thus the normalisation becomes $\sum_a A_{a|x}=\Pi$. The above implies 
\begin{align}\label{props}
&\Pi^2=\Pi, \quad \tr(\Pi)\leq r, \quad (\Pi\otimes \openone) \rho_\text{AB}=\rho_\text{AB},\quad \Pi A_{a|x}=A_{a|x}.
\end{align}

We now construct a hierarchy of SDP relaxations that give outer approximations of the set $\mathcal{S}_r$. We assume that Alice's measurements are projective, i.e.~that $ A_{a|x}A_{a'|x}=\delta_{a,a'}A_{a|x}$, but we will later show that also general measurements can be considered with this method. To this end, we first define the set, $L$, of all operators relevant in the steering scenario, namely the state $\rho_\text{AB}$, the relevant support space $\Pi$ and the measurements $\{A_{a|x}\}$; $L=\{\rho_\text{AB},\Pi\otimes \openone_d,\{A_{a|x}\otimes \openone_d\}_{a,x}\}$. Next, we consider some set $S\supset L$ whose elements are  monomials over $L$.  We use this to define a block-matrix whose block-rows and block-columns are indexed by $u,v\in S$,
\begin{align}\label{moment}
	& \Gamma=\sum_{u,v}\ketbra{u}{v}\otimes \Gamma_{u,v}, \quad \text{with} \quad \Gamma_{u,v}=\tr_\text{A}\left(u^\dagger v\right).
\end{align}
The reason that this is a useful definition is two-fold.~(i) $\Gamma$ is positive semidefinite  by construction \cite{a_footnote}, which is crucial for an SDP formulation. (ii) The relevant  constraints for the steering scenario, namely the assemblage, the Schmidt number of $\rho_\text{AB}$ and the orthogonality of the measurements, can be imposed as (SDP-compatible) linear constraints on $\Gamma$. This is achieved by fixing specific blocks of $\Gamma$, in particular
\begin{align}\label{eq:constraints_gamma}
		& \Gamma_{A_{a|x},\rho}=\sigma_{a|x}, \qquad \Gamma_{\Pi,\Pi}\leq r\openone, \qquad \Gamma_{A_{a|x},A_{a'|x}}=0,
\end{align}
for $a\neq a'$. Thus, the existence of a positive semidefinite $\Gamma$ subject to linear constraints of the above type is a necessary condition for $\bm\sigma \in \mathcal{S}_r$. This problem is therefore an SDP. Importantly,  neither the number of variables nor the size of $\Gamma$ depends on the Schmidt number; $r$ enters only as a linear constraint, as in \eqref{eq:constraints_gamma}. Furthermore, while we have so far assumed that $\rho_\text{AB}$ is pure, note that the set of feasible $\Gamma$ is convex. Therefore, the SDP criterion holds also for any convex combination of pure states with Schmidt number at most $r$, i.e.~it applies also to mixed states.  

Choosing the monomials included in $S$ is a degree of freedom in our construction. By extending $S$ with more monomials, the relaxations of $\mathcal{S}_r$ become increasingly precise.  A standard choice of $S$ is to consider a sequence $S_k$ for $k=1,2,3,\ldots$, where $S_k$ contains all monomials of length at most $k$. Thus,  $S_1\subset S_2\subset\ldots \subset S_\infty$ and it therefore correspond to a hierarchy of SDPs in which each subsequent step is at least as constraining as the previous. Note however that one does not need to select $S$ to be of this form. An often attractive compromise between accuracy and computational cost is to select $S_k$ and extend it with an incomplete set of monomials of length $k+1$. These are called intermediate levels.

It is instructive to consider an example of how the method applies to an illustrative choice of monomials \cite{c_footnote},  $S=\{\Pi, \rho, A_{0|0}, A_{0|1}, A_{1|0}\rho\}$ . Using the relations \eqref{props} and also projective measurements $\Gamma$ becomes
\begin{equation*}
	\includegraphics[width=0.7\columnwidth]{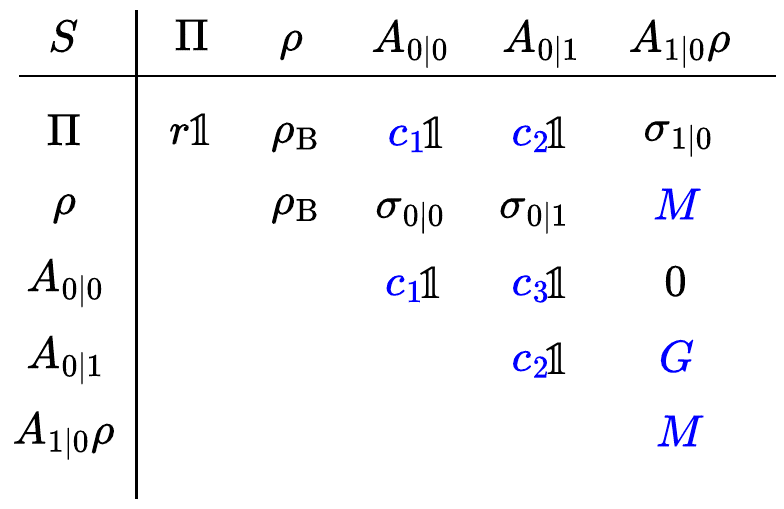}
\end{equation*}
where we have omitted the lower-triangle blocks due to symmetry and optimally chosen equality in Eq.~\eqref{eq:constraints_gamma} for $\Gamma_{\Pi,\Pi}$. The {\color{black}\textbf{black}} entries are fixed by \eqref{eq:constraints_gamma} whereas the {\color{blue}\textbf{blue}} entries are free variables in the SDP. Specifically, $\blue{c_1}, \blue{c_2}$ and $\blue{c_3}$ are non-negative scalars, $\blue{M}$ is a Hermitian matrix and $\blue{G}$ is any matrix. The first follows from  $\Gamma_{\Pi,A_{a|x}}=\tr(\Pi A_{a|x})\openone=\tr(A_{a|x})\openone$, where $\tr(A_{a|x})$ is some non-negative number. The second is due to $\Gamma_{\rho,A_{1|0}\rho}=\tr_\text{A}(\rho (A_{1|0}\otimes \openone) \rho)$ being Hermitian but otherwise unknown. The third is due to  $\Gamma_{A_{0|1},A_{1|0}\rho}$ being an unknown matrix. Thus, if the SDP returns that there exists no choice of the free variables such that $\Gamma\succeq 0$, we conclude that $\bm\sigma\notin \mathcal{S}_r$. Notice that the SDP method  can also be applied to bound witnesses. For this, one needs only to assign $\sigma_{a|x}$ as an unknown assemblage in the block $\Gamma_{\rho,A_{a|x}}=\sigma_{a|x}$. Since any witness \eqref{witness} is a linear functional over these blocks, $W$ can be used as the objective function of the SDP optimised over domain $\Gamma\succeq 0$. 

Furthermore, the SDP can be adapted to relax the assumption of projective measurements. For this, we use Naimark's theorem to represent an $m$-outcome non-projective measurement on the space $\Pi$ as a projective measurement on an extended Hilbert  space of dimension $rm$ \cite{Oszmaniec2017}. This means that $A_{a|x}$ no longer is supported only on $\Pi$ but instead on $\openone_{rm}$. Hence, the final condition in Eq.~\eqref{props} is replaced with $\openone_{rm}A_{a|x}=A_{a|x}$. To account for this in the SDP, we need only to add $\openone_{rm}$ to the list $L$, implying e.g.~$\Gamma_{\mathds{1},\mathds{1}}=rm\openone$. Nevertheless, we note that  non-projective measurements rarely have advantages over projective ones in steering \cite{Bavaresco2017, Nguyen2018, Zhang2024, Renner2024}.

\begin{figure}
	\centering
	\includegraphics[width=1\columnwidth]{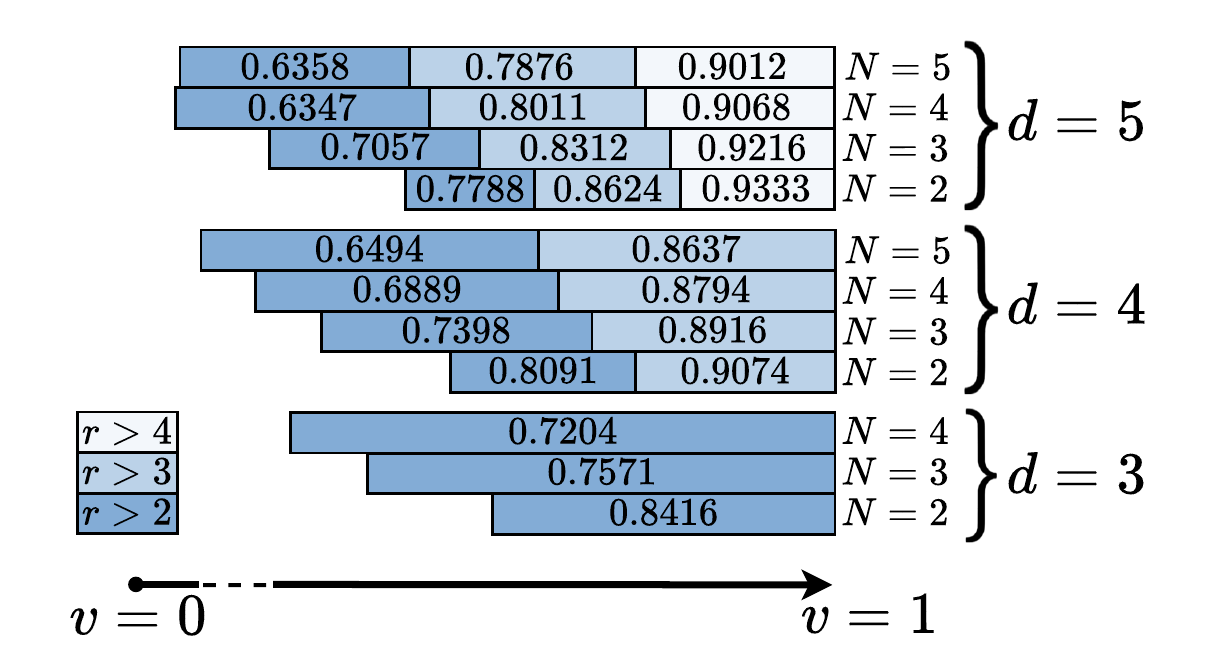}
	\caption{Upper bound on the critical visibility for genuine $r$-dimensional steering under projective measurements of a $d$-dimensional isotropic state under $N$ MUBs. 
		All results were computed on a standard laptop with 32GB RAM using the YALMIP package \cite{Lofberg2004}.}\label{Fig:res_steering}
\end{figure}

\textit{Case studies for the Schmidt number.---} We now address the central question: does the SDP hierarchy perform well in practice for detecting genuine high-dimensional steering? We examine this through case studies.

Consider a noisy maximally entangled state, $\rho_v=v\ketbra{\phi^+_d}+\frac{1-v}{d^2}\mathbbm{1}$, where $v\in[0,1]$ is the visibility.  If Alice performs  basis measurements $\{A_{a|x}\}$, the corresponding assemblage takes the form $\sigma_{a|x}=\frac{v}{d}(A_{a|x})^T+\frac{1-v}{d^2}\mathbbm{1}$.
Let Alice's measurements be a set of $N$ mutually unbiased bases (MUBs), defined by $\tr(A_{a|x}A_{a'|x'})=\frac{1}{d}$ for any pair $x\neq x'$.
These bases are common in experiments and they have also been the focus of previous literature \cite{SebSteering2021, Designolle2022, CompleteHierarchy}. We first use the SDP hierarchy with projective measurements to determine upper bounds on the critical value of $v$ above which the assemblage certifies Schmidt number $r$. For this, we have introduced $v$ as a variable in our SDP relaxation and the objective is to minimise it such the SDP is feasible. For $d=3,4,5$, we have considered all non-trivial values of $r$ for different choices of $N$. The visibility bounds  are given in Fig.~\ref{Fig:res_steering} and the specific relaxation levels used in these computations are detailed in \cite{supp_mat}. Contrasting previous SDP methods \cite{CompleteHierarchy}, we are easily able to go beyond the lowest Schmidt numbers using only a standard laptop. To further showcase the scalability of the method, we have considered dimensions $d=7,8$ for $N=2$ and evaluated the SDP relaxation \cite{b_footnote} for all values of $r$; see complete results in \cite{supp_mat}. For example, in the most demanding case of $(d,r)=(8,7)$ we obtain $v\leq 0.970$.

We now go further and show that the method also gives relevant and scalable results when permitting general measurements. To this end, we have considered the certification of the maximal Schmidt number ($r>d-1$) in the above scenario of $N=2$ MUBs in dimensions $d=3,4,5,6$. We obtain
\begin{table}[h!]
	\begin{tabular}{|l|*{4}{c|}}\hline
		$d$ & $3$ & $4$ & $5$ & $6$ \\ \hline
		$v$ & $ 0.8537 $ & $ 0.9196 $ & $ 0.9443$ & $ 0.9574 $ \\ \hline
	\end{tabular}.
\end{table}

\noindent The relaxation levels are given in \cite{supp_mat}. These results are only somewhat weaker than those obtained under projective measurements in Fig.~\ref{Fig:res_steering} and they improve on the best known results in the literature \cite{SebSteering2021, Designolle2022, CompleteHierarchy} in all cases except the simplest one ($d=3$).

Furthermore, we can also consider  assemblages obtained with measurements other than MUBs. For instance, we can apply to $\rho_v$ a set of $d^2$ binary measurements corresponding to a symmetric informationally complete set of projections. These are given by $\{P_x,\openone-P_x\}_{x=1}^{d^2}$
with $P_x$ being a rank-one projector with the property that $\Tr(P_iP_j)=\frac{1}{d+1}$ for $i\neq j$. We have considered the cases of $d=3$ and $d=4$. For $d=3$, we find genuine three-dimensional steering under general measurements for $v>0.7906$. For $d=4$, we certify a maximal Schmidt number ($r>3$)	 when  $v>0.8634$ \cite{d_footnote}. 

\textit{Hierarchy for entanglement fidelity.---} We now turn to the problem of bounding the lowest entanglement fidelity $F(\rho_\text{AB})$ compatible with a given $\bm\sigma$. For this, note that the entanglement fidelity \eqref{fidelity} can be expressed as an SDP by using the Choi representation of a quantum channels \cite{cavalcanti2023}. This alternative form reads $\{F(\rho_\text{AB})=\max \tr(\rho_\text{AB}X_{\Lambda^\dagger})|X_{\Lambda^\dagger}\succeq 0, \tr_\text{B}(X_{\Lambda^\dagger})=\frac{\mathbbm{1}}{d}\}$ . From the strong duality theorem of SDP, this can equally well be written as the minimisation \cite{supp_mat}
\begin{equation}\label{fidelitydual}
	\begin{aligned}
		F(\rho_\text{AB})=& \min_Z & &  \frac{1}{d}\tr(Z)\\
		& \text{s.t.} & & Z\otimes \mathbbm{1} -\rho_{\text{AB}}\succeq0 \ .
	\end{aligned}
\end{equation}

We use the representation \eqref{fidelitydual} to construct a hierarchy of SDPs for lower-bounding  $F(\rho_\text{AB})$ from $\bm\sigma$. Similar to before, we select $L=\{\openone\otimes \openone,\{A_{a|x}\otimes \openone\}_{a,x}\}$ and let $S$ be a set of monomials over $L$. This time, however, we make use of two separate SDP matrices,  $\Upsilon^{(w)}=\sum_{u,v\in S}\ketbra{u}{v}\otimes \Upsilon_{u,v}^{(w)}$ labeled by $w=\rho$ and $w=Z$ respectively. The reason for these labels is that we define their $d\times d$ blocks as  $\Upsilon_{u,v}^{(\rho)}=\tr_\text{A}\left(u^\dagger \rho v\right)$ and $\Upsilon_{u,v}^{(Z)}=\tr_\text{A}\left(u^\dagger (Z\otimes \openone) v\right)$ respectively. These definitions are employed for reasons similar to the previous case of Schmidt numbers; they permit us to impose the relevant constraints from the steering problem on the SDP relaxation. To see this, note first that $\Upsilon^{(\rho)}$ is positive semidefinite by construction. Secondly, the assemblage appears in its blocks; $\Upsilon^{(\rho)}_{A_{a|x},\mathds{1}}=\sigma_{a|x}$. Thirdly, the entanglement fidelity constraint can be imposed in an SDP-compatible way thanks to $\Upsilon^{(Z)}$. To that end, we first recover the objective in \eqref{fidelitydual} as the trace over  $\Upsilon^{(Z)}_{\mathds{1},\mathds{1}}$ because $f\equiv\frac{1}{d}\tr(Z)=\frac{1}{d^2}\tr(\Upsilon^{(Z)}_{\mathds{1},\mathds{1}})$. Next, we can also write the second line in \eqref{fidelitydual} as a semidefinite constraint in our relaxation by imposing $\Upsilon^{(Z)}-\Upsilon^{(\rho)}\succeq 0$. Thus, we have obtained an SDP relaxation: minimise  $f$ when $\Upsilon^{(\rho)}$ and $\Upsilon^{(Z)}$ satisfy the above semidefinite constraints. This is  guaranteed to return a lower bound on $F(\rho_\text{AB})$ for any choice of monomials $S$. In supplementary \cite{supp_mat} we provide an illustrative example of how this method can be used for a simple choice of $S$.

Alternatively,  we can also estimate the entanglement fidelity from a witness \eqref{witness} instead of from an assemblage. This needs only a small modification:  treat instead  $\Upsilon^{(\rho)}_{A_{a|x},\mathds{1}}=\sigma_{a|x}$ as an unknown assemblage and write the witness as $W=\sum_{a,b,x,y} c_{abxy}\tr(\Upsilon^{(\rho)}_{A_{a|x},\mathbbm{1}}B_{b|y})$. The SDP relaxation amounts to minimising $f$ for $W$ constrained to given value.

\textit{Case studies for entanglement fidelity.---} We apply the SDP relaxation hierarchy to estimate the entanglement fidelity based on a well-known witness for high-dimensional steering \cite{Skrzypczyk2015, Marciniak2015}, and test it by fixing Bob's measurements to be $N$ $d$-dimensional MUBs, $\{B_{b|y}\}$. Specifically, we take as a witness the bound on the probability of Alice and Bob having identical outcomes when selecting the same classical input, i.e.
\begin{equation}\label{ineq}
W\!=\!\frac{1}{N}\sum_{k=1}^N\sum_{j=1}^d\Tr\left[(A_{j|k}\otimes B_{j|k})\rho_{\text{AB}}\right]\!\leq \!\frac{1}{N}\left(1+\frac{N-1}{\sqrt{d}}\right). \nonumber 
\end{equation}
The inequality holds for all separable states and measurements in Alice's side. This inequality is maximally violated with $W=1$ if  Alice and Bob share the maximally entangled state $\rho_{\text{AB}}=\ketbra{\phi^+_d}$ and  select $A_{j|k}=B_{j|k}^T$.

\begin{figure}[t!]
	\centering
	\includegraphics[width=\columnwidth]{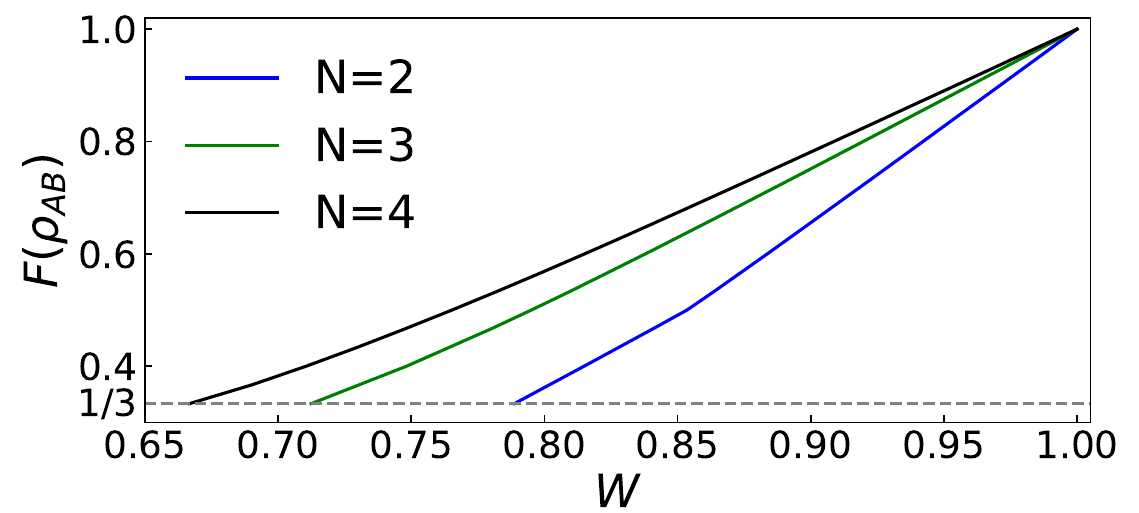}
	\caption{Lower bound on one-sided device-independent entanglement fidelity as function of the MUB-witness $W$. Results are computed for $d=3$ and $N=2,3,4$ MUBs.}
	\label{Fig:res_fidelity}
\end{figure}

When the steering inequality is violated, we evaluate the SDP relaxation to bound from below the entanglement fidelity of the state. Specifically, we have considered the case of $d=3$ with any number of MUBs, namely $N=2,3,4$. Our selected hierarchy levels correspond to SDP matrices of sizes $183$, $295$ and $261$ respectively and are specified in \cite{supp_mat}. The results are illustrated in Fig.~\ref{Fig:res_fidelity}. We obtain a non-trivial entanglement fidelity ($F>1/3$) whenever the steering inequality is violated. To investigate the accuracy of the bounds, we use a seesaw optimization \cite{Tavakoli2024} to also obtain upper bounds on the entanglement fidelity. Details on this method are given in \cite{supp_mat}. Even though this method is not guaranteed to converge, it matches our lower bound for $N=2$ up to numerical precision. For $N=3$ and $N=4$ the maximum discrepancy between the upper and lower bound remains small, specifically at $ 6\times 10^{-3}$ and $2\times 10^{-2}$ respectively.  

\textit{Conclusions.--} We have  introduced hierarchies of semidefinite programs to bound the set of quantum correlations realisable in steering experiments both when (i) the Schmidt number of the source is restricted, and (ii) when the fidelity of the source with the maximally entangled $d$-dimensional state is restricted. We demonstrated through  case studies that the former method both leads to relevant bounds in high-dimensional steering scenarios and that it allows for computations that are scalable in the Schmidt number of the state. Similarly, we showed that the latter method can lead to optimal fidelity bounds for standard high-dimensional steering witnesses. Both methods can readily be applied to data in past and future high-dimensional steering experiments. Our implementation of the SDP hierarchies is available at \cite{github_rep}.

An open problem is whether  SDP  hierarchies converge to the relevant sets in the limit of large relaxation level. A natural further endeavour is to investigate whether the use of symmetries (see e.g.~\cite{Tavakoli2019, Nguyen2020, Ioannou2022}) can be leveraged to reduce the computational cost of our SDPs significantly, mainly w.r.t the number of bases measured by Alice and/or the physical dimension considered.  

Finally, it is interesting to note that certification of the Schmidt number in steering experiments has a one-to-one relationship with a dimension-based generalised notion of joint measurability \cite{Jones2023}. This notion asks whether the statistics of a measurement can be fully reproduced by first compressing a state to a lower dimension and then performing lower-dimensional measurements on it. Our results of Schmidt number certification can therefore also be reinterpreted as criteria for this form of joint measurability.

\begin{acknowledgments}
We thank S\'ebastien Designolle and Roope Uola for discussions. This work is supported by the Wenner-Gren Foundations, by the Knut and Alice Wallenberg Foundation through the Wallenberg Center for Quantum Technology (WACQT) and the Swedish Research Council under Contract No.~2023-03498.
\end{acknowledgments}
	
\onecolumngrid
\newpage
\appendix

\section{Additional details on SDP relaxations}\label{AppGamma}
In Table \ref{Tab:App_lvl} we give the relaxation levels used to compute the visibility bounds of high-dimensional steering with MUBs given in the main text. For given pair $(d,N)$, the same relaxation level was used for the different values of $r$.

\begin{table}[h!]
	\begin{tabular}{|l|*{4}{c|}}\hline
		\diagbox{$d$}{$N$}
		& $2$ & $3$ & $4$ & $5$ \\ \hline
		$3$ & $3$ & $3$ & $2+AAA^*$ & - \\ \hline
		$4$ & $3$ & $2+\rho AA+A_{0|0}AA$ & $2$ & $1+\rho A+AA$ \\\hline
		$5$ & $2+AAA$ & $2$ & $2$ & $1+\rho A+AA^*$ \\\hline
	\end{tabular}
	\caption{Relaxation levels employed in the SDP used to upper bound the critical visibility for genuine $r$-dimensional steering of a $d$-dimensional isotropic state under $N$ MUBs. The intermediate levels are defined by adding selected sets of operators to a complete relaxation level. To simplify the notation of intermediate levels, we have defined $AA^*$ as the set of ordered combination of $A_{a|x}A_{a'|x'}$ with $ax<a'x'$, where $ax$ represents the concatenation of the outcome $a$ and the input $x$ as a two digit number. Furthermore, we define $A_{a|0}$ and $A_{0|x}$ as the list of measurements with fixed input(output) $0$ and output(input) varying from $0$ to $d(N)$.}
	\label{Tab:App_lvl}
\end{table}

For the special case of critical visibilities certifying that $r>2$, we have also employed an additional simplification. Since the support space of Alice is two-dimensional, at most two of the $d$ projectors appearing in her POVM can be non-zero. This implies additional constraints on the moment matrix elements. To give an example consider $\tr(A_{0|0}A_{0|1}A_{1|0}A_{1|1})$. In general, this element would be a complex variable in our SDP. Using the aforementioned property, it is however possible to rewrite it as
\begin{equation}
	\tr(A_{0|0}A_{0|1}A_{1|0}A_{1|1})=\tr(A_{0|0}(\Pi-A_{0|1}^\perp)A_{1|0}A_{1|1})=-\tr(A_{0|0}A_{0|1}^\perp A_{1|0}A_{1|1})
\end{equation}
Where we defined the non zero projector othogonal to $A_{0|1}$ as $A_{0|1}^\perp$. The last trace is either zero when $A_{1|1}\neq A_{0|1}^\perp$ or a real valued element when $A_{1|1} = A_{0|1}^\perp$. We can therefore define it as a real variable.

In the main text we mention computing bounds on the critical visibility of the isotropic state in dimensions $d=7,8$ using $N=2$ MUBs. For each $r$, the results are given in Table~\ref{Tab78}.

\begin{table}[h!]
	\begin{tabular}{|l|*{6}{c|}}\hline
		\diagbox{$d$}{$r$}
		& $2$ & $3$ & $4$ & $5$ & $6$ & $7$ \\ \hline
		$7$ & $0.7531$ & $0.8247$ & $0.8797$ & $0.9253$ & $0.9649$ & - \\\hline
		$8$ & $0.7385$ & $0.8056$ & $0.8571$ & $0.9000$  & $0.9371$ & $0.9701$ \\ \hline
	\end{tabular}
	\caption{Upper bounds on the critical visibility for genuine $r$-dimensional steering of a isotropic state for $d=7,8$ with  $N=2$ MUBs.}\label{Tab78}
\end{table}
When certifying the maximal Schmidt number without the assumption of projective measurements on Alice's side, we employed level $3+AAA$, $2+\rho AA+AAA$, $2+AAA$ and $2+\Pi AA$ for dimensions $3,4,5,6$.

Finally, the SDP relaxation levels used to compute the entanglement fidelity bounds in the main text for the MUB-witness for $N=2,3,4$ MUBs in dimension $d=3$ are as follows. For $N=2$, we use level 4. For $N=3$, we use level $2+A_{a|0}AA$. For $N=3$ we use level $2+A_{0|0}AA$. 

All our selections of levels are motivated by compromises between accuracy of the relaxation and computational viability. That is, for every case study, we have strived to select a monomial list, $S$, that is large enough to imply constraints that are strong enough to obtain interesting bounds, but small enough to be computationally reasonable on a standard laptop computer.

\section{Dual SDP formulation of maximally entangled fraction}\label{AppDual}
The maximally entangled fraction of a state is commonly defined as
\begin{align}
	F_{\Phi^+}(\rho_\text{AB})= & \ \max_U \ \ \bra{\Phi^+}(U\otimes\mathds{1})\rho_\text{AB}(U^\dagger\otimes\mathds{1})\ket{\Phi^+},
\end{align}
where $U$ is a unitary that Alice can use to locally align her state with the maximally entangled state $\ket{\Phi^+}$. Note that without loss of generality, one can consider a unitary only on Alice's side since any unitary permitted on Bob's side can be absorbed into Alice's unitary via the relation $(O\otimes\mathds{1})\ket{\Phi^+}=(\mathds{1}\otimes O^T)\ket{\Phi^+}$. We go one step further and permit Alice to apply a general CPTP map, $\Lambda$, to her local state. This can be viewed as an extraction channel over which the optimisation is evaluated,
\begin{align}\label{eq:fidelity_lambda}
	& F(\rho_\text{AB})=\  \max_\Lambda \ \ \bra{\Phi^+}(\Lambda\otimes\mathds{1})[\rho_\text{AB}]	\ket{\Phi^+}.
\end{align}
We refer to the maximally entangled fraction $F(\rho_\text{AB})$ as the entanglement fidelity. It can conveniently be recast in terms of the Choi state of the adjoint channel. 
\begin{equation}
	\begin{split}
		\bra{\Phi^+}(\Lambda\otimes\mathds{1})[\rho_\text{AB}]\ket{\Phi^+}&=\tr\left[\left(\Lambda\otimes\mathds{1}\right)[\rho_\text{AB}]\ket{\Phi^+}\bra{\Phi^+}\right]\\
		&=\tr\left[\rho_\text{AB}\left(\Lambda^\dagger\otimes\mathds{1}\right)[\ket{\Phi^+}\bra{\Phi^+}]\right]\\
		&=\tr\left[\rho_\text{AB}X_{\Lambda^\dagger}\right].
	\end{split}
\end{equation}
In the second equality we used the definition of the adjoint channel and in the last equality we have introduced the Choi state  $X_{\Lambda^\dagger}=\left(\Lambda^\dagger\otimes\mathds{1}\right)[\ket{\Phi^+}\bra{\Phi^+}]$. The adjoint channel is completely positive and unital. Hence, the Choi state is characterised by the constraints $X_{\Lambda^\dagger}\succeq 0$ and $\tr_B\left[X_{\Lambda^\dagger}\right]=\dfrac{1}{d}\mathds{1}$. This allows us to evaluate the entanglement fidelity as the following SDP
\begin{equation}\label{eq:app_primal}
	\begin{aligned}
		F(\rho_\text{AB})=\ & \max_{X_{\Lambda^\dagger}} & &  \tr[\rho_\text{AB}X_{\Lambda^\dagger}]\\
		& \text{s.t.} & &  \tr_B\left[X_{\Lambda^\dagger}\right]=\dfrac{1}{d}\mathds{1}\\
		& & & X_{\Lambda^\dagger}\succeq 0
	\end{aligned}
\end{equation}
For the purpose of using our moment matrix relaxation method, we are interested in expressing this as a minimisation problem instead. Therefore, we consider the dual formulation. To find the dual, we minimise the Lagrangian 
\begin{equation}
	\mathcal{L}(X_{\Lambda^\dagger},Z,\Xi)=	\tr[\rho_\text{AB}X_{\Lambda^\dagger}]+\tr\left[Z\left( \dfrac{1}{d}\mathds{1}-\tr_BX_{\Lambda^\dagger}\right)\right]+\tr[\Xi X_{\Lambda^\dagger}],
\end{equation}
where  $Z\succeq 0$ and $\Xi\succeq0$. The Lagrangian can be simlified to
\begin{equation}\label{eq:lagrangian}
	\begin{split}
		\mathcal{L}(X_{\Lambda^\dagger},Z,\Xi)&=\tr[\rho_\text{AB}X_{\Lambda^\dagger}]+\dfrac{1}{d}\tr[Z]-\tr\left[(Z\otimes\mathds{1})X_{\Lambda^\dagger}\right]+\tr[\Xi X_{\Lambda^\dagger}]\\
		&=\dfrac{1}{d}\tr[Z]+\tr\left[X_{\Lambda^\dagger}\left(\rho_\text{AB}-(Z\otimes\mathds{1})+\Xi\right)\right]
	\end{split}
\end{equation}
To minimise the Lagrangian over the dual variables we require that
\begin{equation}
	\rho_\text{AB}+\Xi=Z\otimes\mathds{1}
\end{equation}
Since $\Xi\succeq0$ we can write the final minimisation as the SDP 
\begin{equation}\label{eq1}
	\begin{aligned}
		F(\rho_\text{AB})=\ & \min & &  \dfrac{1}{d}\tr(Z)\\
		& \text{s.t.} & &  Z\otimes \mathds{1} \succeq \rho_\text{AB}.
	\end{aligned}
\end{equation}

\section{Seesaw optimization}\label{AppSeesaw}
We optimise the value of the witness over the measurements of Alice and the shared state subject to a limited entanglement fidelity, $f$. This optimisation is written
\begin{equation}\label{eq:app_seesaw}
	\begin{aligned}
		W(f)=\ & \max_{A,\rho,Z} & &  W\\
		& \text{s.t.} & & \dfrac{1}{d}\tr(Z)=f\\
		& & &  Z\otimes \mathds{1} - \rho_\text{AB} \succeq 0.
	\end{aligned}
\end{equation}
This is a nonlinear problem since the witness is quadratic in the optimisation variables
\begin{equation}
	W=\sum_{a,x}\tr\left[(A_{a|x}\otimes B_{a|x})\rho_\text{AB}\right].
\end{equation}
The nonlinear optimization problem in Eq \ref{eq:app_seesaw}, can be efficiently approximated by alternate linear optimizations. In detail, fixing Alice's measurements $A_{a|x}$ the problem is linear in the shared state $\rho_\text{AB}$ and the dual variable $Z$ and vice versa. It means that we can efficiently solve by the means of a semidefinite program the two separate problems:
\begin{minipage}{0.45\textwidth}
	\begin{equation}\label{eq:app_seesaw_1}
		\begin{aligned}
			& \max_{\rho, Z} & &  W_f(\rho, \tilde{A})\\
			& \text{s.t.} & & \dfrac{1}{d}\tr(Z)=f\\
			& & &  Z\otimes \mathds{1} - \rho_\text{AB} \succeq 0
		\end{aligned}
	\end{equation}
\end{minipage}
\hfill
\begin{minipage}{0.45\textwidth}
	\begin{equation}\label{eq:app_seesaw_2}
		\begin{aligned}
			& \max_{A, Z} & &  W_f(\tilde{\rho}, A)\\
			& \text{s.t.} & & \dfrac{1}{d}\tr(Z)=f\\
			& & &  Z\otimes \mathds{1} - \rho_\text{AB} \succeq 0
		\end{aligned}
	\end{equation}
\end{minipage}\\
\par\medskip
Our algorithm then is as follows:
\begin{itemize}
	\item Chose random assignments $A_{a|x}=\tilde{A}_{a|x}$
	\item Solve the SDP \ref{eq:app_seesaw_1} optimising over all states $\rho_\text{AB}$ and matrices $Z$ and save the solutions $\tilde{\rho}_{AB}$ and $\tilde{Z}$ 
	\item Solve the SDP \ref{eq:app_seesaw_2} optimising over all measurements $A_{a|x}$ and matrices $Z$ and save the solutions $\rho_\text{AB}$ and $\tilde{Z}$
	\item Repeat step 2 and 3 until convergence
\end{itemize}
Must be noted that to perform the aforementioned optimization the dimension of the uncharacterised party has to be fixed. In our application we impose $d_A=2d^2$.  

\section{Illustrative example of SDP relaxation method for entanglement fidelity}\label{GammaFidelity}

In this section we give an instructive example of how to work with the SDP realxation method for bounding the entanglement fidelity. Our example is based on a simple choice of monomials to illustrate how the relevant steering constraints are imposed in the SDP matrix formalism. To this end, let us consider bounding the entanglement fidelity for the MUB-witness in dimension $d=3$ and using $N=2$ bases $\{B_{b|y}\}$. Recall that this witness reads
\begin{equation}
	W\!=\!\frac{1}{N}\sum_{k=0}^{N-1}\sum_{j=0}^{d-1}\tr\left[(A_{j|k}\otimes B_{j|k})\rho_{\text{AB}}\right]\!\leq \!\frac{1}{N}\left(1+\frac{N-1}{\sqrt{d}}\right).
\end{equation}
We will select the first level of the hierarchy, i.e.~we choose $S=\{\openone, A_{0|0}, A_{0|1}, A_{1|0}, A_{1|1}\}$. Note that we do not need to include the third outcome $A_{2|x}$ because it is fixed by normalisation, $A_{2|x}=\openone-A_{0|x}-A_{1|x}$.

Using the definition of the moment matrices:
\begin{align}
	&\Upsilon^{(\rho)}=\sum_{u,v\in S}\ketbra{u}{v}\otimes \Upsilon_{u,v}^{(\rho)}, \qquad \Upsilon_{u,v}^{(\rho)}=\tr_\text{A}\left(u^\dagger \rho v\right)\\
	&\Upsilon^{(Z)}=\sum_{u,v\in S}\ketbra{u}{v}\otimes \Upsilon_{u,v}^{(Z)}, \qquad \Upsilon_{u,v}^{(Z)}=\tr_\text{A}\left(u^\dagger Z\otimes \openone v\right),
\end{align}
we get\\
\begin{minipage}{0.9\textwidth}
	\begin{equation*}
		\includegraphics[width=1.\columnwidth]{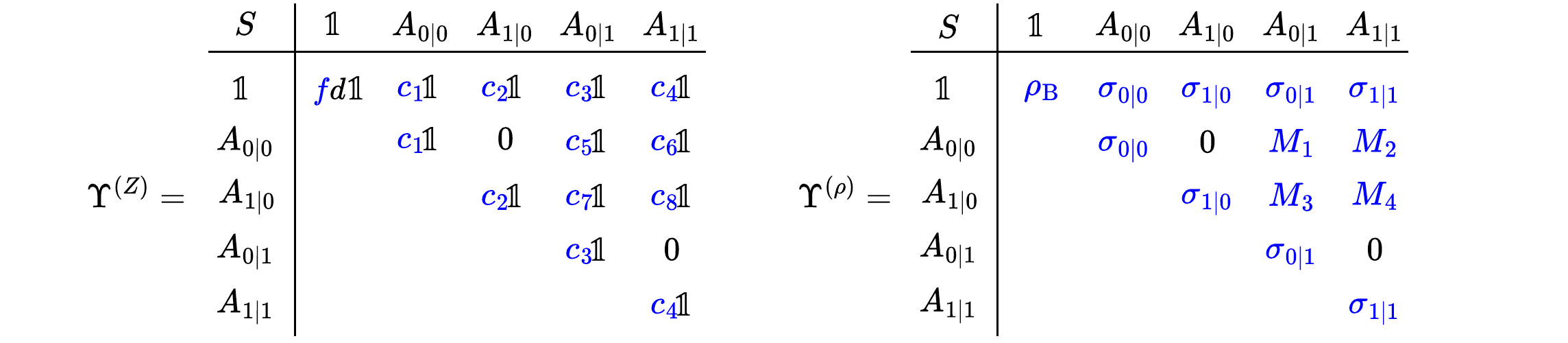}
	\end{equation*}
\end{minipage}\hfill
\begin{minipage}{0.1\textwidth}
	\begin{equation}\label{eq:upsilons}
		\
	\end{equation}
\end{minipage} \par\medskip
In \eqref{eq:upsilons}, the \textbf{black} entries represents entries that can be fixed exploiting  the steering conditions. In the above, we have $\Upsilon^{(Z)}_{A_{a|x}A_{a'|x}}=0$ and $\Upsilon^{(\rho)}_{A_{a|x},A_{a'|x}}=0$ for $a\neq a'$ due to orthogonality of the projective measurements.  Also, the constrained value of the witness enters as a linear functional constraint over suitable blocks of $\Upsilon^{(\rho)}$, specifically
\begin{equation}
	W=\frac{1}{3}\sum_{k=0,1}\sum_{j=0,1,2} \tr(\Upsilon^{(\rho)}_{A_{j|k},\mathbbm{1}}B_{j|k}),
\end{equation} 
where we use the normalisation of the measurements to write $\Upsilon^{(\rho)}_{A_{2|k},\mathds{1}}=\Upsilon^{(\rho)}_{\mathds{1},\mathds{1}}-\Upsilon^{(\rho)}_{A_{0|k},\mathds{1}}-\Upsilon^{(\rho)}_{A_{1|k},\mathds{1}}$.

The \blue{\textbf{blue}} entries in \eqref{eq:upsilons} are free variables in the SDP. Specifically,
\begin{itemize}
	\item we have $\Upsilon^{(Z)}_{\mathds{1},\mathds{1}}=\tr_\text{A}{(Z\otimes\openone)}=\blue{f}d\openone$, with $\blue{f}$ representing the entanglement fidelity and $d$ the dimension. This is the variable we aim to minimise in the fidelity estimation problem.
	\item $\blue{c_1}, \blue{c_2}, \blue{c_3}$ and $\blue{c_4}$ are non-negative scalars. This follows from $\Upsilon^{(Z)}_{\mathds{1},A_{a|x}}=\tr(ZA_{a|x})\openone$, where $\tr(ZA_{a|x})$ is non-negative since for projective measurements $\tr(ZA_{a|x})=\tr(A_{a|x}ZA_{a|x})$ holds and $A_{a|x}ZA_{a|x}\succeq 0$.
	\item $\blue{c_{5}}, \blue{c_{6}}, \blue{c_{7}}$ and $\blue{c_{8}}$ are complex numbers, this is due to $\Upsilon^{(Z)}_{A_{a|x},A_{a'|x'}}=\tr(ZA_{a'|x'}A_{a|x})\openone$ being unknown when $x\neq x'$
	\item $\blue{\sigma_{a|x}}$ are Hermitian semidefinite positive matrices representing Bob's assemblage. This comes directly from $\Upsilon^{(\rho)}_{A_{a|x},\mathds{1}}=\tr_\text{A}(\rho(A_{a|x}\otimes\openone))=\sigma_{a|x}$. The assemblage respects the no-signaling condition $\sum_a \blue{\sigma_{a|x}}=\blue{\rho_{\text{B}}}$, with $\tr\blue{\rho_{\text{B}}}=1$. Note that we consider $\sigma_{a|x}$ as a free variable only when estimating the fidelity from a witness parameter. In contrast, if the fidelity is estimated directly from a given assemblage, $\sigma_{a|x}$ would be a fixed block in the matrix (i.e.~coloured black). 
	\item $\blue{M_i}$ are arbitrary complex matrices. This is because $\Upsilon^{(\rho)}_{A_{a|x},A_{a'|x'}}=\tr_\text{A}(\rho(A_{a'|x'}A_{a|x}\otimes\openone))$ when $x\neq x'$.
\end{itemize}

The SDP relaxation for bounding the lowest value of the entanglement fidelity $F(\rho)$ compatible with the witness value $W$ is then written  
\begin{equation}\label{eq:app_seesaw_2}
	\begin{aligned}
		F(\rho)\geq  & \min_{\Upsilon^{(\rho)},\Upsilon^{(Z)}} & & \frac{1}{d^2}\tr(\Upsilon^{(Z)}_{\mathds{1},\mathds{1}})\\
		&\quad \text{s.t.} & &  W=\frac{1}{3}\sum_{k=0,1}\sum_{j=0,1,2} \tr(\Upsilon^{(\rho)}_{A_{j|k},\mathbbm{1}}B_{j|k})\\
		& & & \Upsilon^{(\rho)}\succeq 0 \\
		& & & \Upsilon^{(Z)}-\Upsilon^{(\rho)}\succeq 0. 
	\end{aligned}
\end{equation}
Recall that the constraint $\Upsilon^{(\rho)}\succeq 0$ holds by definition of $\Upsilon^{(\rho)}$ and that $\Upsilon^{(Z)}-\Upsilon^{(\rho)}\succeq 0$ serves as a semidefinite relaxation of the constraint appearing in the dual formulation of the entanglement fidelity \eqref{eq1}. 

		\bibliography{bibliography}

\begin{thebibliography}{47}
\bibitem [{a_f()}]{a_footnote}%
  \BibitemOpen
  \href@noop {} {}\bibinfo {howpublished} {This can be shown by noticing that
  $\forall \ket{\alpha}$ it holds that
  $\bracket{\alpha}{\Gamma}{\alpha}=\tr\left(QQ^\dagger\right)\geq 0$ where
  $Q=\sum_{u\in S} (\openone_{A} \otimes \bra{\alpha}_{BC})(u_{AB}^\dagger
  \otimes \ket{u}_{C})$.}\BibitemShut {Stop}%
\bibitem [{c_f()}]{c_footnote}%
  \BibitemOpen
  \href@noop {} {}\bibinfo {howpublished} {Here we have for the sake of a clean
  presentation written $\Pi$ and $A_{a|x}$ instead of $\Pi\otimes \openone$ and
  $A_{a|x}\otimes \openone$.}\BibitemShut {Stop}%
\bibitem [{b_f()}]{b_footnote}%
  \BibitemOpen
  \href@noop {} {}\bibinfo {howpublished} {For dimension $7$, level 2 of the
  hierarchy was employed. For dimension $8$ the intermediate level $1+\rho
  A+A_{a|0}A$ was used.}\BibitemShut {Stop}%
\bibitem [{d_f()}]{d_footnote}%
  \BibitemOpen
  \href@noop {} {\bibinfo {title} {These results were obtained using hierarchy
  level 2.}}\BibitemShut {Stop}%
\end{thebibliography}

\end{document}